\documentclass[journal]{IEEEtran}%[12pt, draftclsnofoot, onecolumn]

\ifCLASSINFOpdf

\else

\fi

\usepackage{mathrsfs}
\usepackage{amsmath}
\usepackage{amsfonts}
\usepackage{amsthm}
\usepackage{amssymb}
\usepackage{graphicx}
\usepackage{subfigure}
\usepackage{indentfirst}
\usepackage{array}
\usepackage{cite}
\usepackage{enumerate}
\usepackage{bm}
\usepackage[linesnumbered,ruled,vlined]{algorithm2e}
\usepackage{algorithmic}
\usepackage{multirow}
\usepackage{epstopdf}
\usepackage{verbatim}
\usepackage{stfloats}
\usepackage{color}
\usepackage{booktabs}
\usepackage{footnote}

\newtheorem{lemma}{\bf{Lemma}}

\newtheorem{myDef}{Definition}
\SetKwComment{Comment}{//}{}

\begin{document}

\title{\huge{Sphere Constraint based Enumeration Methods to Analyze the Minimum Weight Distribution of Polar Codes}}

\author{Jinnan~Piao,~\IEEEmembership{Student Member,~IEEE}, Kai~Niu,~\IEEEmembership{Member,~IEEE}, Jincheng~Dai,~\IEEEmembership{Member,~IEEE}, and Chao~Dong,~\IEEEmembership{Member,~IEEE}
\thanks{This work is supported by National Key R\&D Program of China (No. 2018YFE0205501), the National Natural Science Foundation of China (No. 61671080), China Post-Doctoral Science Foundation (No. 2019M660032) and BUPT Excellent Ph.D. Students Foundation (No. CX2019218).}
\thanks{The authors are with the Key Laboratory of Universal Wireless Communications, Ministry of Education, Beijing University of Posts and Telecommunications (BUPT), Beijing 100876, China (email: piaojinnan@bupt.edu.cn, niukai@bupt.edu.cn, daijincheng@bupt.edu.cn,  dongchao@bupt.edu.cn).}
\vspace{-1em}
}

\maketitle

\begin{abstract}

In this paper, the minimum weight distributions (MWDs) of polar codes and concatenated polar codes are exactly enumerated according to the distance property of codewords. We first propose a sphere constraint based enumeration method (SCEM) to analyze the MWD of polar codes with moderate complexity. The SCEM exploits the distance property that all the codewords with the identical Hamming weight are distributed on a spherical shell. Then, based on the SCEM and the Plotkin's construction of polar codes, a sphere constraint based recursive enumeration method (SCREM) is proposed to recursively calculate the MWD with a lower complexity. Finally, we propose a parity-check SCEM (PC-SCEM) to analyze the MWD of concatenated polar codes by introducing the parity-check equations of outer codes. Moreover, due to the distance property of codewords, the proposed three methods can exactly enumerate all the codewords belonging to the MWD. The enumeration results show that the SCREM can enumerate the MWD of polar codes with code length up to $2^{14}$ and the PC-SCEM can be used to optimize CRC-polar concatenated codes.
\end{abstract}


\IEEEpeerreviewmaketitle

\section{Introduction}\label{section_introduction}

\IEEEPARstart{P}{olar} codes have been proved to achieve the capacity by the successive cancellation (SC) decoding as the code length goes to infinity \cite{arikan}. However, when the code length is small or medium, the performance is unsatisfying. Thus, successive cancellation list (SCL) decoding \cite{talvardyscl, niuscl} and successive cancellation stack decoding \cite{SCS} are introduced to improve the performance of polar codes. Furthermore, the performance is improved by the CRC-aided SCL (CA-SCL) decoding \cite{niu_CASCL} which introduces the CRC detector into the SCL decoding. Thanks to its excellent performance, polar codes have been adopted as the coding scheme for the control channel of the enhanced Mobile Broadband (eMBB) service category in the fifth generation wireless communication systems (5G) \cite{3GPP_5G_polar, FAR_DL}.

The weight distribution of codewords is the distance spectrum of polar codes, which can be used to evaluate the maximum-likelihood (ML) performance \cite{ADSCL}. However, enumerating the distance spectrum has exponential complexity and it is almost impossible for long code length. In the high signal-to-noise ratio (SNR) region, the minimum weight distribution (MWD) is the main factor influencing the ML performance \cite{LinShu}. Thus, the ML performance can be evaluated by MWD instead of the distance spectrum.

To analyze the MWD of polar codes,
%\cite{dmin} proves that the minimum Hamming weight of polar codes is the minimum row weight of the generator matrix.
Li \emph{et al.} \cite{ADSCL} propose an SCL method with excessively large list size to enumerate codewords and analyze MWD. However, due to the large consumption of memory and high complexity, implementing this method on a memory-constrained computer is difficult.
Thus, the hard disk is used in \cite{dsliu} to reduce the number of survival paths and decrease the consumption of memory and a multi-level SCL method is proposed to reduce the list size in \cite{CRCdesign}. Nevertheless, these SCL based methods still have high complexities.

Besides, concatenated polar codes \cite{SCPC}, especially CRC-polar concatenated codes \cite{niu_CASCL}, have better error performance than polar codes, since the distance spectrum of polar codes is improved by resorting to concatenated schemes.
An uniform interleaver approach \cite{CRCds} is proposed to analyze the distance properties of concatenated polar code ensembles, but it cannot obtain the distance spectrum of CRC-polar concatenated codes with definite CRC polynomial. In addition, since this approach enumerates all the codewords, its complexity is extremely high.

In this paper, we exploit the distance property of codewords to exactly evaluate the MWDs of polar codes and concatenated polar codes.
The distance property is that the codewords with the identical Hamming weight are distributed on a spherical shell.
Hence, a sphere constraint with the minimum Hamming weight can early prune a large amount of unnecessary codewords for analyzing MWD. In addition, the sphere constraint ensures that all the codewords with the minimum weight could be enumerated exactly.

A sphere constraint based enumeration method (SCEM) is first proposed to analyze the MWD of polar codes with moderate complexity.
The process of SCEM similar to that of the sphere decoding (SD) algorithm \cite{SD, niu_SD, efficient_SD, CASD} is regarded as a depth-first tree search, which has negligible memory overhead compared with the SCL based methods. In the SCEM, the sphere constraint with the minimum Hamming weight is used to evaluate the MWD. Thus, the paths satisfying the sphere constraint in the search tree are reserved and the MWD is evaluated exactly. In comparison, the SCL based methods cannot guarantee to enumerate all the codewords with the minimum Hamming weight when the list size is small or medium. Then, although the paths violating the constraint are early pruned to reduce the redundant search, the complexity of SCEM is still too high to evaluate the MWD of long polar codes. Therefore, a sphere constraint based recursive enumeration method (SCREM) is proposed to analyze the MWD with lower complexity on the basis of the SCEM and the Plotkin's construction. Additionally, inspired by the CRC-aided SD (CA-SD) algorithm \cite{CASD}, a parity-check SCEM (PC-SCEM) is proposed to analyze the MWD of concatenated polar codes by introducing the parity-check equations of outer codes.

The main contributions of this paper are summarized as follows:
\begin{enumerate}[1)]
    \item We first propose the SCEM to exactly enumerate all the codewords belonging to the MWD by exploits the distance property that all the codewords with the identical Hamming weight are distributed on a spherical shell.
    \item The SCREM is proposed to recursively analyze the MWD of polar codes with lower complexity compared with the SCEM. In the SCREM, we first prove the property that the MWD of a polar code is related with the MWD of the two component polar codes in terms of the Plotkin's construction. Based on the property, the MWD of the polar code is directly decided without search when the minimum Hamming weight of the two component codes is identical and the complexity of enumerating the MWD of the polar code can be efficiently reduced when the two component codes have different minimum Hamming weight.
    \item The PC-SCEM is proposed to exactly analyze the MWD of concatenated polar codes by introducing the parity-check equations of outer codes. The parity-check equations are utilized to ensure that all the codewords enumerated are valid codewords. Then, to match the search order of PC-SCEM, Gaussian elimination is used to transform the parity-check equations into new forms. Due to the newly parity-check equations and the sphere constraint, all the codewords belonging to the MWD of concatenated polar codes are exactly enumerated.
\end{enumerate}

The experimental results show that the proposed SCEM and SCREM with code length 128 have up to $10^4$ and $10^8$ times lower complexity compared with the SCL methods, respectively.
The MWD analysis results show that the SCREM can enumerate the MWD of polar codes with code length up to $2^{14}$ and the PC-SCEM can analyze the MWD of CRC-polar concatenated codes to optimize the CRC polynomial.

%The analysis results show the complexity of the proposed methods is up to $10^4$ times lower than the SCL based methods. Moreover, the distance spectrum of polar codes with code length $2^{14}$ can be obtained by the RSS method accurately. Then, the CRC polynomial of CRC-polar concatenated codes can be optimized according to the distance spectrum obtained by the PC-SS method.

The remainder of the paper is organized as follows. Section II describes the preliminaries of polar codes, SD algorithm and distance spectrum. In Section III, the distance property of codewords and the SCEM are described. The SCREM is provided to recursively analyze the MWD in terms of the Plotkin's construction in Section IV.
Section V presents the PC-SCEM to evaluate the MWD of concatenated polar codes.
The MWD and the complexity evaluation are provided in Section VI. Section VII concludes this paper.

\section{Notations and Preliminaries}\label{section_model}

\subsection{Notation Conventions}

In this paper, the lowercase letters, e.g., $x$, are used to denote scalars. The bold lowercase letters (e.g., ${\bf{x}}$) are used to denote vectors. Notation ${{\bf x}_i^j}$ denotes the subvector $(x_i,\cdots,x_j)$ and $x_i$ denotes the $i$-th element of ${\bf{x}}$. The sets are denoted by calligraphic characters, e.g., $\cal{X}$, and the notation $|\cal{X}|$ denotes the cardinality of $\cal{X}$. In addition, ${\cal X} \backslash x$ denotes the set with element $x$ excluded.
The bold capital letters, such as $\bf{X}$, are used to denote matrices.
%The notation $\bf{X}^\prime$ stand for the transpose transpose of $\mathbf{X}$.
The element in the $i$-th row and the $j$-th column of matrix $\bf{X}$ and the $i$-th row of matrix $\bf{X}$ are written as $x_{i,j}$ and ${\bf X}_{i}$, respectively.
Furthermore, we write ${\bf{F}}^{\otimes n}$ to denote the $n$-th Kronecker power of $\bf{F}$ and the bit-reversal permutation is denoted by $\pi(\cdot)$.
Throughout this paper, $\bf 0$ and $\bf 1$ mean an all-zero vector and an all-one vector, respectively.

\subsection{Polar Codes and Concatenated Polar Codes}

Polar codes depend on the polarization effect \cite{arikan} of the matrix
${{\bf F} = \begin{bmatrix}
\begin{smallmatrix}
1&0\\
1&1
\end{smallmatrix}
\end{bmatrix}}$.
For an $(N,K)$ polar code with code length $N = 2^n$ and code rate $R = K/N$, the polarization effect generates $N$ polarization subchannels.
Each subchannel has different reliability and the information bits are transmitted in the $K$ most reliable subchannels. Therefore, the information set of polar codes defined by ${\cal A}$ with cardinality $|{\cal A}|=K$ is composed of the indices of the $K$ most reliable subchannels and it is a subset of the index set $\left\{1,2,\cdots,N\right\}$. Then, the frozen set ${\cal A}^c$ with cardinality $|{\cal A}^c|=N-K$ is a complementary set of ${\cal A}$.
The codeword ${\bf c}$ of polar codes is calculated by ${\bf c}  = {\bf u}{\bf B}{\bf G} = {\bf v}{\bf G}$, where ${\bf u}$ is an $N$-length information sequence, ${\bf B}$ is a bit-reversal permutation matrix, ${\bf G}$ is ${\bf F}^{\otimes n}$ and ${\bf v} = {\bf u}{\bf B}$.
The information sequence ${\bf u}$ is generated by assigning $u_i$ to information bit if $i \in {\cal A}$, and assigning $u_i$ to $0$ if $i \in {\cal A}^c$.
Then, according to ${\bf v} = {\bf u}{\bf B}$, an another information set ${\cal B} = \left\{j|j=\pi(i - 1)+1, i \in {\cal A}\right\}$ is obtained, which means $v_j$ is an information bit if $j \in {\cal B}$. Here, $\pi(\cdot)$ is a bit-reversal permutation.

For an $(N,K_I)$ concatenated polar code, the inner code is an $(N,K)$ polar code and the outer code is a $(K,K_I)$ binary linear block code. The message sequence $\bf b$ is first encoded by the binary linear block code to obtain the encoded sequence $\bf s$. Then, $\bf s$ is treated as the information bits of an $(N,K)$ polar code and it is inserted into the information sequence $\bf u$ in terms of the information set $\cal A$. Furthermore, a codeword of the concatenated polar code is calculated as ${\bf c}  = {\bf u}{\bf B}{\bf G} = {\bf v}{\bf G}$.

Without loss of generality, the binary-input additive white Gaussian noise (BI-AWGN) channel and BPSK modulation are considered in this paper. Thus, each coded bit $c_i \in \left\{0,1\right\}$ is modulated into the transmitted signal by $x_i = 1 - 2c_i$. Then, the received sequence is ${\bf y} = {\bf x} + {\bf n}$, where $n_i$ is i.i.d. AWGN with zero mean and variance $\sigma^2$.

\subsection{Sphere Decoding Algorithm}

ML decoding of polar codes is equivalent to the following minimization problem
\begin{equation}\label{min_problem}
\begin{aligned}
{\hat {\bf{v}}} = \mathop {\arg \min }\limits_{{\bf{x}} } {\left\| {{\bf{y}} - {\bf{x}}} \right\|^2} = \mathop {\arg \min }\limits_{{\bf{v}} } {\left\| {{\bf y} - \left({\bf 1} - 2{\bf v}{\bf G}\right)} \right\|^2},
\end{aligned}
\end{equation}
where ${\bf 1}$ is an all-one vector of length $N$.
SD algorithm can solve the problem by enumerating the possible sequence $\bf v$ satisfying the sphere constraint
\begin{equation}\label{SD_euclidean_distance}
  m\left({\bf v}_1^N\right) \triangleq {\left\| {{\bf y} - \left({\bf 1} - 2{\bf v}{\bf G}\right)} \right\|^2} \le r^2,
\end{equation}
where $r$ denotes the radius for the SD search and $m\left({\bf v}_1^N\right)$ is the squared Euclidean distance along with the sequence ${\bf v}_1^N$. Noting that ${\bf G}$ is a lower triangular matrix, we can define the partial squared Euclidean distance as
\begin{equation}\label{SD_partial_euclidean_distance}
  m\left({\bf v}_i^N\right) \triangleq \sum\limits_{k = i}^N {\left| {{y_k} - \left(1-2\cdot{\mathop  \oplus \limits_{j=k}^N \left(v_jg_{j,k}  \right)}\right)} \right|}^2,
\end{equation}
which can be computed recursively as
\begin{equation}\label{SD_partial_euclidean_distance_recursive}
    m\left({\bf v}_i^N\right) = m\left({\bf v}_{i+1}^N\right) + {\left| {{y_i} - \left(1-2\cdot{\mathop  \oplus \limits_{j=i}^N \left(v_jg_{j,i}  \right)}\right)} \right|}^2,
\end{equation}
where ${\bf v}_i^N$ denotes the bit decisions from the $i$-th bit to the $N$-th bit, and `$\oplus$' denotes summation over $GF(2)$. According to (\ref{SD_partial_euclidean_distance_recursive}), the SD algorithm can be regarded as a depth-first search on the tree and the search order is from the $N$-th bit $v_N$ to the first bit $v_1$. Once a valid sequence $\bf v$ satisfying the sphere constraint is found, the radius is updated by $\sqrt{m\left({\bf v}_i^N\right)}$.
To find the ML decoding sequence, SD adaptively updates the radius $r$. In this process, $r$ decreases rapidly so that the ML solution is efficiently captured.

\subsection{Distance Spectrum}

The distance spectrum of an $(N,K)$ binary linear block code, designated by $A_d$, is the number of codewords of the code with the Hamming weight $d$. The pairwise error probability between two codewords modulated by BPSK differing in $d$ positions and coherently detected in the AWGN channel is $Q\left(\sqrt{\frac{{2dRE_b}}{{{N_0}}}}\right)$, where $E_b$ is the energy of the transmitted bit, $N_0$ is the one-sided power spectral density of AWGN and
\begin{equation}\label{Q_function}
Q(x)=\frac{1}{{\sqrt {2\pi }  }}\int_x^\infty  {{e^{ - \frac{{{t^2}}}{2}}}dt}
\end{equation}
is the probability that a random Gaussian variable with zero mean and unit variance exceeds the value $x$. We assume that an all-zero codeword $\bf 0$ is transmitted to analyze the ML performance. The union bound of ML decoding performance can be written as
\begin{equation}\label{union_bound}
P_e
\le \sum\limits_{d = {d_{\min }}}^N {{A_d}Q\left( {\sqrt {\frac{{2dR{E_b}}}{{{N_0}}}} } \right)} .
\end{equation}

Then, since the MWD (i.e., $d_{\min}$ and $A_{d_{\min}}$) is the main factor influencing the ML performance when the $E_b/N_0$ is large, (\ref{union_bound}) can be approximated as
\begin{equation}\label{union_bound_a}
P_e \approx {{A_{d_{\min}}}Q\left( {\sqrt {\frac{{2{d_{\min}}R{E_b}}}{{{N_0}}}} } \right)},
\end{equation}
where $d_{\min}$ is the minimum Hamming weight of the linear block code and ${A_{d_{\min}}}$ is the number of the codewords with $d_{\min}$. In this paper, the approximate union bound (AUB) calculated by (\ref{union_bound_a}) is used to evaluate the performance of polar codes.

\section{Sphere Constraint based Enumeration Method}

In this section, we first illustrate the codewords distribution of polar codes and the idea of SCEM. Then, the detailed description of SCEM is provided on the basis of the codewords distribution.

\subsection{An Outline of SCEM}

\begin{figure}[t]
\setlength{\abovecaptionskip}{0.cm}
\setlength{\belowcaptionskip}{-0.cm}
  \centering{\includegraphics[scale=1.15]{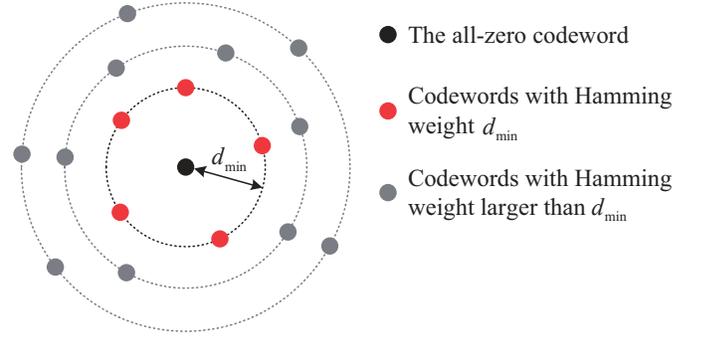}}
  \caption{The description of the codewords distribution with the minimum Hamming weight $d_{\min}$ in the codeword space.}\label{SS_Structure}
\end{figure}
\begin{figure}[t]
\setlength{\abovecaptionskip}{0.cm}
\setlength{\belowcaptionskip}{-0.cm}
  \centering{\includegraphics[scale=0.97]{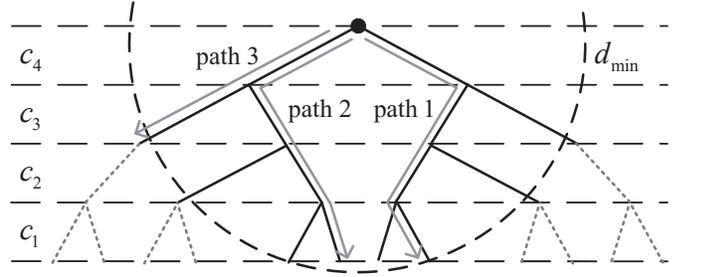}}
  \caption{The illustration of a binary search tree with code length $4$ and sphere constraint $d_{\min}$.}\label{Tree_Structure}
\end{figure}

The codewords distribution of polar codes is illustrated in Fig. \ref{SS_Structure}.
For a polar code, the codewords with the identical Hamming weight are distributed on a spherical shell. Then, in order to analyze the MWD, the number of all the codewords with the minimum Hamming weight needs to be counted. Based on the codewords distribution, these codewords are covered by a sphere whose radius is the minimum Hamming weight. Thus, a method enumerating all these codewords constrained by the sphere can evaluate the MWD exactly.

SD can find the closest decoded sequence from the received sequence in codeword space under the radius constraint. Inspired by this idea, we propose SCEM to enumerate the codewords under sphere constraint and analyze the MWD. Similar to the SD, SCEM is regarded as a depth-first tree search as well.
However, since SCEM is just used to enumerate the codewords, the noise is unnecessary.

Fig. \ref{Tree_Structure} is a toy example to illustrate the process of SCEM.
A binary search tree with code length $4$ under the sphere constraint $d_{\min}$ is described in Fig. \ref{Tree_Structure}.
The branches of the $i$-th level in the tree are associated with $c_{N-i+1}$. Each path from the root node to a leaf node represents a codeword. In Fig. \ref{Tree_Structure}, path 1 and path 2 satisfy the sphere constraint and the two paths are reserved in the search tree.
On the contrary, path 3 violates the sphere constraint. Thus, all the paths attached to path 3 are pruned from the search tree, since the corresponding codewords are out of the sphere. Therefore, the proposed SCEM reserves all the codewords satisfying the sphere constraint to analyze the MWD and prunes unnecessary codewords to reduce the redundant search.

In comparison, since the SCL based methods are breadth-first tree search, a lot of paths need to be stored in the memory, which results in large memory overhead. Moreover, due to no constraint used in the SCL based methods, the unnecessary search is unavoidable. Furthermore, some codewords with the minimum Hamming weight may be lost in the SCL based methods when the list size is not enough.

\begin{algorithm}[t]
\setlength{\abovecaptionskip}{0.cm}
\setlength{\belowcaptionskip}{-0.cm}
\caption{The SCEM method: $({\cal{T}}, A_{d_{\min}}) = {\sf SCEM} \left(N, K, {\cal B}\right)$}\label{SS}

\KwIn {The code length $N$, the information bit length $K$ and the information set $\cal B$;}
\KwOut {$\cal T$ is a set composed of the codewords with $d_{\min}$ and $A_{d_{\min}}$ is the number of the codewords with $d_{\min}$;}
Initialize $d_{\min} \leftarrow \mathop {\min }\limits_{i \in {\cal B}} \left(wt\left({\bf G}_{i}\right)\right)$, ${\cal T} \leftarrow \varnothing$ and $A_{d_{\min}} \leftarrow 0$\;
Initialize  the index of searching bit $i \leftarrow N$\;
Initialize ${\bf v} \leftarrow {\bf 0}$ and ${\bf c} \leftarrow {\bf 0}$\;

\While{$i \le N$}
{
    \If(\tcp*[f]{information bit}){$i \in {\cal B}$}
    {
        %\Comment{information bit}
        ${v}_i \leftarrow \mathop {\arg \min }\limits_{{ v}_i \in \{0,1\}} \left(c_i\right)$ and $c_i \leftarrow 0$;
    }
    \Else(\tcp*[f]{frozen bit})
    {
        %\Comment{frozen bit}
        ${v}_i \leftarrow 0$ and $c_i \leftarrow {\left({\mathop  \oplus \limits_{j=i}^N \left(v_jg_{j,i}  \right)}\right)}$;
    }
    \If(\tcp*[f]{satisfy the sphere constraint}){$d\left({\bf v}_{i}^N\right) \le d_{\min}$}
    {
        %\Comment{satisfy the sphere constraint}
        \If{$i > 1$}
        {
            $i \leftarrow i  - 1$\;
        }
        \Else
        {
            ${\cal T} \leftarrow {\cal T} \cup \{{\bf c}\}$ and             $A_{d_{\min}} \leftarrow A_{d_{\min}} + 1$\;
            Go to Step 16\;
        }
    }
    \Else(\tcp*[f]{prune the search tree})
    {
        %\Comment{prune the search tree}
        \While{$i \le N$}
        {
            \If{$i \in {\cal B}$ and $c_i = 0$}
            {
                $v_i \leftarrow v_i \oplus 1$ and $c_i \leftarrow 1$\;
                 Go to Step 9\;
            }
            \Else
            {
                $i \leftarrow i + 1$\;
            }
        }
    }
}
${\cal T} \leftarrow {\cal T} - \{{\bf 0}\}$ and $A_{d_{\min}} \leftarrow A_{d_{\min}} - 1$\;

\end{algorithm}

\subsection{Detailed Description of SCEM}

For an $(N,K)$ polar code $\cal C$, all the codewords with the Hamming weight $d_{\min}$ in the codeword space are on the surface of a sphere with radius $d_{\min}$. Based on this, SCEM is proposed to enumerate all these codewords and analyze the MWD.
The whole procedure is described in Algorithm \ref{SS}.

The Hamming weight of a codeword $\bf c$ is denoted as $wt({\bf c}) = \sum\nolimits_{i = 1}^N {{c_i}} $. Then, according to \cite{CRCdesign}, the minimum Hamming weight of polar codes is the minimum row weight of generator matrix, i.e.,
\begin{equation}\label{minimum_Hamming_weight}
d_{\min} = \mathop {\min }\limits_{i \in {\cal B}} \left(wt\left({\bf G}_{i}\right)\right).
\end{equation}
According to (\ref{minimum_Hamming_weight}), the sphere constraint used in the SCEM is decided, which is
\begin{equation}\label{distance_sphere_search}
wt\left({\bf c}\right) \le d_{\min}.
\end{equation}
Thus, the codewords satisfying (\ref{distance_sphere_search}) is on the surface of the sphere constraint except ${\bf c} = {\bf 0}$.

Then, since ${\bf G}$ is a lower triangular matrix, $c_i$ is only related to the subvector ${\bf v}_i^N$, i.e.,
\begin{equation}\label{c_i}
c_i = {\mathop  \oplus \limits_{j=i}^N \left(v_jg_{j,i}  \right)}.
\end{equation}
Thus, the partial Hamming weight of $\bf c$ is defined as
\begin{equation}\label{partial_distance_sphere_search}
d\left({\bf v}_i^N\right) \triangleq wt\left({\bf c}_i^N\right)=\sum\limits_{k = i}^N {\left({\mathop  \oplus \limits_{j=k}^N \left(v_jg_{j,k}  \right)}\right)},
\end{equation}
which can be calculated recursively as
\begin{equation}\label{partial_distance_sphere_search_recursive}
d\left({\bf v}_i^N\right) = d\left({\bf v}_{i+1}^N\right) + {\left({\mathop  \oplus \limits_{j=i}^N \left(v_jg_{j,i}  \right)}\right)}.
\end{equation}

According to (\ref{partial_distance_sphere_search_recursive}),
the process of enumerating all the codewords on the surface of the sphere can be treated as a depth-first tree search and the search order is from $v_N$ to $v_1$.
Then, when $v_i$ is decided, $d\left({\bf v}_i^N\right)$ is decided as well. Therefore, the sphere constraint (\ref{distance_sphere_search}) can be simplified as
\begin{equation}\label{distance_sphere_search_simplified}
d\left({\bf v}_i^N\right) \le d_{\min},
\end{equation}
which means that the Hamming weight of the codewords attached to ${\bf c}_i^N$ is larger than $d_{\min}$ when (\ref{distance_sphere_search_simplified}) is false. Hence, these codewords need to be pruned from the search tree to avoid the useless search.

Algorithm \ref{SS} describes the entire procedure of SCEM, where $N$ is the code length, $K$ is the information bit length and $\cal B$ is the information set about $\bf v$. Without loss of generality, for describing the SCEM easily, we set the branch with $c_i = 0$ as the first searching branch during deciding information bit $v_i$. Then, when the search of the branch with $c_i = 0$ is completed, SCEM continues to search the branch with $c_i = 1$.

In Algorithm \ref{SS}, the search order is from $v_N$ to $v_1$. Thus, in terms of (\ref{partial_distance_sphere_search_recursive}), when a bit $v_i$ is decided, whether $d\left({\bf v}_{i}^N\right)$ satisfies the Hamming weight constraint (\ref{distance_sphere_search_simplified}) or not is judged.
If satisfying, the search continues to decide next bit $v_{i-1}$ until that a codeword with Hamming weight $d_{\min}$ is enumerated. If not, the nodes attached to the path ${\bf c}_{i}^N$ are pruned from the search tree and the search goes on from a new branch of the tree. By repeating the search process, all the $A_{d_{\min}}$ codewords with Hamming weight $d_{\min}$ is enumerated and these codewords are recorded into a codeword set ${\cal T}$. Thus, the MWD ${\cal T}$ is obtained by Algorithm \ref{SS}, i.e.,
\begin{equation}\label{Tset}
{\cal T} = \left\{{\bf c}|wt({\bf c}) = d_{\min}, {\bf c} \in {\cal C}\right\}.
\end{equation}

\section{Sphere Constraint based Recursive Enumeration Method}

In this section, we first prove the MWD relationship between a polar code and two component polar codes based on the Plotkin's construction. Then, according to the MWD relationship and the SCEM, we design the SCREM.

\subsection{MWD Relationship Based on the Plotkin's Construction}

According to the Plotkin's construction of polar codes \cite{Plotkin}, a polar code can be divided into two component polar codes. Then, we find that the MWD of the polar code is related with the MWDs of the two component codes and prove the MWD relationship. Based on this, the MWD can be enumerated recursively.

To prove the MWD relationship, we first describe the Plotkin's construction of polar codes as follows. For an $(N, K)$ polar code $\cal C$ with information set ${\cal B}$, the encoding process can be expressed as
\begin{equation}\label{Plotkin¡¯s construction}
\begin{aligned}
{\bf c} &= {\bf v}{\bf G} \\
&= \left({\bf v}', {\bf v}''\right)\left[ {\begin{array}{*{20}{c}}
{\bf G}'&0\\
{\bf G}'&{\bf G}'
\end{array}} \right] \\
&= \left({\bf c}' \oplus {\bf c}'',{\bf c}''\right),
\end{aligned}
\end{equation}
where ${\bf G}'$ is ${\bf F}^{\otimes (n-1)}$, ${\bf c}'$ is ${\bf v}'{\bf G}'$ and ${\bf c}''$ is ${\bf v}''{\bf G}'$. Then, ${\bf c}'$ and ${\bf c}''$  are the codewords of an $(\frac{N}{2}, K')$ polar code ${\cal C}'$ and an $(\frac{N}{2}, K'')$ polar code ${\cal C}''$, respectively. The information set of ${\cal C}'$ is denoted by ${\cal B}'$ and
\begin{equation}\label{B'}
{\cal B}' = \left\{{i}\left|i \in {\cal B}, 1 \le i \le \frac{N}{2}\right. \right\}.
\end{equation}
Similarly, ${\cal B}''$ is the information set of ${\cal C}''$ and
\begin{equation}\label{B''}
{\cal B}'' = \left\{\left.{i - \frac{N}{2}}\right|i \in {\cal B}, \frac{N}{2} + 1 \le i \le N\right\}.
\end{equation}
In addition, $K' = |{\cal B}'|$ and $K'' = |{\cal B}''|$. The minimum Hamming weight of ${\cal C}'$ and ${\cal C}''$ is denoted by $d'_{\min}$ and $d''_{\min}$, respectively.

Then, in order to prove the MWD relationship among $\cal C$, ${\cal C}'$, and ${\cal C}''$, we first prove Lemma \ref{lemma1_subcode} and Lemma \ref{lemma3_d2_d1} as follows.

\begin{lemma}\label{lemma1_subcode}
${\cal C}'$ is a subcode of ${\cal C}''$.
\end{lemma}
\begin{IEEEproof}
According to the partial order \cite{PO}, if $v_i$ is an information bit, $v_{i + \frac{N}{2}}$ is also an information bit. Thus, if $i \in {\cal B}'$, we have $i \in {\cal B}''$. Therefore, ${\cal C}'$ is a subcode of ${\cal C}''$,
\end{IEEEproof}

\begin{lemma}\label{lemma3_d2_d1}
$d'_{\min}$ and $d''_{\min}$ have three combinations:
\begin{enumerate}[1)]
\item $d'_{\min} = d_{\min}$ and $d''_{\min} = d_{\min}$.
\item $d'_{\min} = d_{\min}$ and $d''_{\min} = \frac{d_{\min}}{2}$.
\item $d'_{\min} > d_{\min}$ and $d''_{\min} = \frac{d_{\min}}{2}$.
\end{enumerate}
\end{lemma}
\begin{IEEEproof}
Due to Lemma \ref{lemma1_subcode}, we have
\begin{equation}\label{d1_larger_d2}
d'_{\min} \ge d''_{\min}.
\end{equation}
According to [16, Sec 4.4], we can obtain
\begin{equation}\label{d_min_plotkin_structure}
d_{\min} = \min\left(2d''_{\min}, d'_{\min}\right).
\end{equation}
Supposing $2d''_{\min} \ge d'_{\min}$ , according to (\ref{d1_larger_d2}) and (\ref{d_min_plotkin_structure}), we have
%$2d''_{\min} \ge d'_{\min} = d_{\min} \ge d''_{\min}.$
\begin{equation}
\left\{ \begin{array}{*{20}{c}}
{2{{d''}_{\min }} \ge {{d'}_{\min }} \ge {{d''}_{\min }}}\\
{{{d'}_{\min }} = {d_{\min }}}
\end{array}\right.
\end{equation}
Then, we obtain
%$\frac{d_{\min}}{2} \le d_2 \le d_{\min}.$
\begin{equation}
\frac{d_{\min}}{2} \le d''_{\min} \le d_{\min}.
\end{equation}
Furthermore, since
\begin{equation}
wt\left({\bf G}_{i}\right) = 2^j, \exists j \in \left\{1,2,\cdots,n \right\},
\end{equation}
we have
\begin{equation}
\left\{ \begin{array}{*{20}{c}}
{{{d'}_{\min }} = {d_{\min }}} ~~~~~~~~~\\
{{{d''}_{\min }} = \frac{{{d_{\min }}}}{2}\;{\rm{or}}~{d_{\min }}}.
\end{array} \right.
\end{equation}

Then, supposing $2d''_{\min} < d'_{\min}$, similarly, according to (\ref{d1_larger_d2}) and (\ref{d_min_plotkin_structure}), we can obtain that %$d_{\min} = 2d''_{\min} \le d'_{\min}.$
$d''_{\min}$ is $\frac{d_{\min}}{2}$ and $d'_{\min} > d_{\min}$.

From the above, Lemma \ref{lemma3_d2_d1} has been proved.
\end{IEEEproof}

According to Lemma \ref{lemma1_subcode} and Lemma \ref{lemma3_d2_d1}, the MWD relationship among $\cal C$, ${\cal C}'$, and ${\cal C}''$ is provided as Lemma \ref{lemma4_distance_spectrum_d1_is_d2}.

\begin{lemma}\label{lemma4_distance_spectrum_d1_is_d2}
Given ${\cal T}$, ${\cal T}'$ and ${\cal T}''$ are the codeword sets of $\cal C$ , ${\cal C}'$ and ${\cal C}''$ with Hamming weights $d_{\min}$, $d'_{\min}$ and $d''_{\min}$, respectively.
    \begin{enumerate}[1)]
        \item When $d'_{\min}=d_{\min}$ and $d''_{\min}=d_{\min}$, we have
            \begin{equation}\label{T1}
            {\cal T} = {\cal T}_1 \cup {\cal T}_2.
            \end{equation}
        \item When $d'_{\min}=d_{\min}$ and $d''_{\min}=\frac{d_{\min}}{2}$, we have
            \begin{equation}\label{T2}
            {\cal T} = {\cal T}_1 \cup {\cal T}_2 \cup {\cal T}_3 \cup {\cal T}_4.
            \end{equation}
        \item When $d'_{\min} > d_{\min}$ and $d''_{\min}=\frac{d_{\min}}{2}$, we have
            \begin{equation}\label{T3}
            {\cal T} = {\cal T}_3.
            \end{equation}
    \end{enumerate}
In (\ref{T1}), (\ref{T2}) and (\ref{T3}),
\begin{equation}
{\cal T}_1 =
\left\{\left({\bf c}', {\bf 0}\right) | {\bf c}' \in {\cal T}'\right\},
\end{equation}
\begin{equation}
{\cal T}_2 = \left\{\left({\bf 0}, {\bf c}'\right) | {\bf c}' \in {\cal T}'\right\},
\end{equation}
\begin{equation}
{\cal T}_3 = \left\{ \left( {\bf c}'',{\bf c}''\right)|{\bf c}''\in{\cal T}'' \right\},
\end{equation}
\begin{equation}
{\cal T}_4 = \{ \left( {\bf c}' \oplus {\bf c}'',{\bf c}''\right)|{\bf c}'\in{\cal T}', {\bf c}''\in{\cal T}'', wt({\bf c}' \oplus {\bf c}'') = \frac{d_{\min}}{2} \}.
\end{equation}
\end{lemma}
\begin{IEEEproof}
See the Appendix.
\end{IEEEproof}
Lemma \ref{lemma4_distance_spectrum_d1_is_d2} describes the relationship among $\cal T$, ${\cal T}'$, and ${\cal T}''$. Based on this, ${\cal T}$ can be directly decided by ${\cal T}'$ and ${\cal T}''$.

\subsection{Detailed Description of SCREM}

\begin{algorithm}[t]
\setlength{\abovecaptionskip}{0.cm}
\setlength{\belowcaptionskip}{-0.cm}
\caption{The SCREM: $({\cal{T}}, A_{d_{\min}}) = {\sf SCREM} \left(N, K, {\cal B}, d_{\min}\right)$}\label{RSS}
\KwIn {The code length $N$, the information bit length $K$, the information set $\cal B$ and the minimum Hamming weight $d_{\min}$;}
\KwOut {$\cal T$ is the codeword set with $d_{\min}$ and $A_{d_{\min}}$ is the number of the codewords with $d_{\min}$;}

Initialize ${\cal T} \leftarrow \varnothing$ and $A_{d_{\min}} \leftarrow 0$ \;
Initialize ${\cal B}'$ and ${\cal B}''$ by (\ref{B'}) and (\ref{B''}), respectively\;
Initialize $K'$ and $K''$ by $|{\cal B}'|$ and $|{\cal B}''|$, respectively\;
$d'_{\min}$ and $d''_{\min}$ are the minimum Hamming weight of ${\cal C}'$ and ${\cal C}''$, respectively\;
\If{$N = 2$ or $K' = 0$ or $K'' = \frac{N}{2}$}
{
    $({\cal{T}}, A_{d_{\min}}) \leftarrow {\sf SCEM} \left(N, K, {\cal B}, d_{\min}\right)$\;
}
\Else
{
    $({\cal T}', A_{d'_{\min}}) \leftarrow {\sf SCREM} \left(\frac{N}{2}, K', {\cal B}', d'_{\min}\right)$\;
    $({\cal T}'', A_{d''_{\min}}) \leftarrow {\sf SCREM} \left(\frac{N}{2}, K'', {\cal B}'', d''_{\min}\right)$\;
    \If{$d'_{\min} = d_{\min}$ and $d''_{\min} = d_{\min}$}
    {
        \Comment{case 1}
        ${\cal T} \leftarrow {\cal T}_1 \cup {\cal T}_2$ and $A_{d_{\min}} \leftarrow 2A_{d'_{\min}}$\;
    }
    \ElseIf{$d'_{\min} = d_{\min}$ and $d''_{\min} = \frac{d_{\min}}{2}$}
    {
        \Comment{case 2}
        Obtain ${\cal T}_4$ by enumerating all the combinations of ${\bf c}'$ and ${\bf c}''$ which satisfy ${\bf c}'\in{\cal T}'$, ${\bf c}''\in{\cal T}''$ and $wt({\bf c}' \oplus {\bf c}'') = \frac{d_{\min}}{2}$\;
        ${\cal T} \leftarrow {\cal T}_1 \cup {\cal T}_2 \cup {\cal T}_3 \cup {\cal T}_4$ and $A_{d_{\min}} \leftarrow 2A_{d'_{\min}} + A_{d''_{\min}} + |{\cal T}_4|$\;
    }
    \ElseIf{$d'_{\min} > d_{\min}$ and $d''_{\min} = \frac{d_{\min}}{2}$}
    {
        \Comment{case 3}
        ${\cal T} \leftarrow {\cal T}_3$ and $A_{d_{\min}} \leftarrow A_{d''_{\min}}$\;
    }
}
\end{algorithm}

In order to exploit Lemma \ref{lemma4_distance_spectrum_d1_is_d2} to analyze the MWD of $\cal C$, the MWD of ${\cal C}'$ and ${\cal C}''$ need to be evaluated first. Fortunately, the SCEM can be used to exactly enumerate the MWD of ${\cal C}'$ and ${\cal C}''$. Then, due to the recursive structure of the generator matrix of polar codes, the MWD of $\cal C$ can be enumerated recursively by using the SCEM and Lemma \ref{lemma4_distance_spectrum_d1_is_d2}.

The SCREM is proposed to enumerate the MWD of polar codes recursively and the method is described as Algorithm \ref{RSS}.
In Algorithm \ref{RSS}, polar code $\cal C$ is first divided into two component polar codes ${\cal C}'$ and ${\cal C}''$ in terms of the Plotkin's construction (step 1 to 4).
Then, $\cal T$ can be obtained by ${\cal T}'$ and ${\cal T}''$ on the basis of Lemma \ref{lemma4_distance_spectrum_d1_is_d2} (step 10 to 16). Also, SCREM is used to enumerate ${\cal T}'$ and ${\cal T}''$ (step 8 and 9). Hence, the MWD of $\cal C$ can be analyzed recursively.
In addition, when $\cal C$ cannot be divided into two component codes, i.e., $N = 2$ or $K' = 0$, or the division cannot reduce the search complexity, i.e., $K'' = \frac{N}{2}$, the recursion is stop and the SCEM is used to enumerate the MWD of $\cal C$ (step 5 and 6).
Thus, according to Lemma \ref{lemma4_distance_spectrum_d1_is_d2} and the SCEM, we can obtain $\cal T$ recursively.

\section{Parity-Check SCEM}

In this section, we first describe the parity-check equations and transform them into new forms to match the search order of the PC-SCEM. Then, the detailed description of PC-SCEM is provided.

\subsection{Parity-Check Equations}

For an $(N,K_I)$ concatenated polar code, the inner code is an $(N,K)$ polar code and the outer code is a $(K,K_I)$ binary linear block code.
The parity-check matrix and the codeword of the binary linear block code is denoted by $\bf H$ and $\bf s$, respectively. Each row of $\bf H$ represents a parity-check equation, i.e.,
\begin{equation}\label{PCE}
\mathop  \oplus \limits_{j = 1}^K {h_{i,j}}{s_j} = 0,~i = 1,2, \cdots ,K_P,
\end{equation}
where $K_P = K - K_I$ is the number of the parity-check equations.

Then, parity-check sets are used in this paper to represent the parity-check equations.
\begin{myDef}\label{d1}
The parity-check sets corresponding to the parity-check equations of $\bf s$ are given as
    \begin{equation}\label{PCS}
        {\cal R}_i({\bf s}) \triangleq \left\{j|h_{i,j} = 1\right\},~i=1,2,\cdots,K_P.
    \end{equation}
\end{myDef}

\begin{algorithm}[t]
\setlength{\abovecaptionskip}{0.cm}
\setlength{\belowcaptionskip}{-0.cm}
\caption{$\left\{{\cal Q}_i\left({\bf v}\right)\right\} = \ ${\tt Transform}$(\left\{{\cal R}_i({\bf v})\right\})$}\label{Qset}
\KwIn {The parity-check sets ${\cal R}_i\left({\bf v}\right), ~l=1,2,\cdots,K_P$;}
\KwOut {Total $K_p$ transformed parity-check sets ${\cal Q}_i\left({\bf v}\right)$\;}
Initialize ${\bf D}$ as a $K_P \times N$ matrix and the $i$-th row of ${\bf D}$ represents the parity-check equation obtained by ${\cal R}_i({\bf v})$\;
Employ GE on the rows of $\bf D$ and obtain a row echelon form matrix $\bf E$\;
Each row of $\bf E$ represents the newly parity-check equation and the corresponding parity-check set is ${\cal Q}_i\left({\bf v}\right)$\;
\end{algorithm}

Since ${\bf s}$ is inserted into ${\bf u}$ in terms of the information set ${\cal A}$, the parity-check sets corresponding to $\bf u$ are defined as
\begin{equation}
  {{\cal R}_i}\left( {\bf{u}} \right) = \left\{ {t\left| {t = f\left( j \right),j \in {{\cal R}_i}\left( {\bf{s}} \right)} \right.} \right\} ,~i=1,2,\cdots,K_P
\end{equation}
where the function $f\left( t \right)$ is the index mapping from $\bf s$ to $\bf u$ and $f\left( t \right)$ is different for various concatenated polar code schemes. Then, the parity-check sets ${{\cal R}_i}\left( {\bf{v}} \right)$ corresponding to $\bf v$ are derived by performing the bit-reversal permutation to all the elements in ${{\cal R}_i}\left( {\bf{u}} \right)$, i.e.,
\begin{equation}
{{\cal R}_i}\left( {\bf{v}} \right) = \left\{k|k=\pi(t-1)+1,t \in {{\cal R}_i}\left( {\bf{u}} \right)\right\}, i=1,2,\cdots,K_P
\end{equation}

In terms of the definition of parity-check sets, for any $i=1,2,\cdots,K_P$, we have
\begin{equation}\label{PCS_equation}
\mathop  \oplus \limits_{j \in {\cal R}_i({\bf s})}s_j = \mathop  \oplus \limits_{t \in {\cal R}_i({\bf u})}u_t =  \mathop  \oplus \limits_{k \in {\cal R}_i({\bf v})}v_k = 0.
\end{equation}

Since the search order of the PC-SCEM method is from $v_N$ to $v_1$, the bit with the least index in each parity-check set ${\cal R}_i \left({\bf v}\right)$ can be directly judged by the previous searched bits. Followed by this, the definition of parity-check bit index of the parity-check set is given.
\begin{myDef}
The index $k_i$ of parity-check bit corresponding to ${\cal R}_i({\bf v})$ is defined as ${k_i} = {\rm min}({\cal R}_i({\bf v}))$.
\end{myDef}

Similar to the CA-SD \cite{CASD}, if two more parity-check sets have the same index of parity-check bit, the \emph{colliding decision} phenomenon will happen where this parity-check bit cannot be uniquely judged. This severe problem leads to the search error.
However, the above colliding decision problem can be solved by the linear combination of multiple parity-check equations.
Thus, to avoid the colliding decision, Gaussian elimination (GE) is used to transform the parity-check equations into new forms to ensure the indices of parity-check bits are different with each other. The process is described in Algorithm \ref{Qset}.

In Algorithm \ref{Qset}, the $i$-th row of a $K_P \times N$ matrix $\bf D$ is first initialized in terms of the parity-check equation obtained by ${\cal R}_i({\bf v})$. Then, GE is used on the rows $\bf D$ to obtain a row echelon form matrix $\bf E$. Finally, the newly parity-check set ${\cal Q}_i\left({\bf v}\right)$ is obtained by the $i$-th row of ${\bf E},i=1,2,\cdots,K_P$ and the parity-check bits of these sets are different due to the row echelon form. Thus, the parity-check bit indices are
\begin{equation}\label{CA-SD_index}
  {k_i} = {\rm min}({\cal Q}_i\left({\bf v}\right)),~i=1,2,\cdots,K_P.
\end{equation}
and these bits can be decided by
\begin{equation}\label{CA-SD_constraint_3}
v_{k_i} = \mathop  \oplus \limits_{k \in ({\cal Q}_i\left({\bf v}\right) \backslash {k_i})}v_k,~i=1,2,\cdots,K_P.
\end{equation}

\begin{algorithm}[t]
\setlength{\abovecaptionskip}{0.cm}
\setlength{\belowcaptionskip}{-0.cm}
\caption{The PC-SCEM}\label{PC-SS}
\KwIn {$N$, $\cal B$ and $\left\{{\cal Q}_i\left({\bf v}\right)\right\}$;}
\KwOut {$\cal T$, $d_{\min}$ and $A_{d_{\min}}$;}
Initialize ${\cal T} \leftarrow \varnothing$, $A_{d_{\min}} \leftarrow 0$ and $r \leftarrow \mathop {\min }\limits_{i \in {\cal B}} \left(wt\left({\bf G}_{i}\right)\right)$\;
Initialize $k \leftarrow N$, ${\bf v} \leftarrow {\bf 0}$ and ${\bf c} \leftarrow {\bf 0}$\;
Initialize ${\cal P} = \left\{k_i|k_i={\rm min}({\cal Q}_i\left({\bf v}\right)),i=1,2,\cdots,K_P\right\}$\;
\While{$A_{d_{\min}} = 0$}
{
    \While{$k \le N$}
    {
        \If(\tcp*[f]{information bit}){$k \in {\cal B} - {\cal P}$}
        {
            %\Comment{information bit}
            ${v}_k \leftarrow \mathop {\arg \min }\limits_{{v}_k \in \{0,1\}} \left(c_k\right)$ and $c_k \leftarrow 0$;
        }
        \ElseIf(\tcp*[f]{parity-check bit}){$k \in {\cal P}$}
        {
            %\Comment{parity-check bit}
            Find $i$ making $k_i = k$\;
            ${v}_k \leftarrow \mathop  \oplus \limits_{t \in ({\cal Q}_i\left({\bf v}\right) \backslash {k_i})}{v}_t$\;
            $c_k \leftarrow {\left({\mathop  \oplus \limits_{j=k}^N \left(v_jg_{j,k}  \right)}\right)}$;
        }
        \Else(\tcp*[f]{frozen bit})
        {
            %\Comment{frozen bit}
            ${v}_k \leftarrow 0$ and $c_k \leftarrow {\left({\mathop  \oplus \limits_{j=k}^N \left(v_jg_{j,k}  \right)}\right)}$;
        }
        \If(\tcp*[f]{satisfy the sphere constraint}){$d\left({\bf v}_{k}^N\right) \le r$}
        {
            %\Comment{satisfy the sphere constraint}
            \If{$k > 1$}
            {
                $k \leftarrow k  - 1$\;
            }
            \Else
            {
                ${\cal T} \leftarrow {\cal T} \cup \{{\bf c}\}$ and             $A_{d_{\min}} \leftarrow A_{d_{\min}} + 1$\;
                Go to Step 21\;
            }
        }
        \Else(\tcp*[f]{prune the search tree})
        {
            %\Comment{prune the search tree}
            \While{$k \le N$}
            {
                \If{$k \in {\cal B} - {\cal P}$ and $c_k = 0$}
                {
                    $v_k \leftarrow v_k \oplus 1$ and $c_k \leftarrow 1$\;
                     Go to Step 14\;
                }
                \Else
                {
                    $k \leftarrow k + 1$\;
                }
            }
        }
    }
    ${\cal T} \leftarrow {\cal T} - \{{\bf 0}\}$ and
    $A_{d_{\min}} \leftarrow A_{d_{\min}} - 1$\;
    \If{$A_{d_{\min}} = 0$}
    {
        $r \leftarrow r+2$\;
    }
    \Else
    {
        $d_{\min} \leftarrow r$\;
    }
}
\end{algorithm}

\subsection{Detailed Description of PC-SCEM}

\begin{table*}[t]
\renewcommand\arraystretch{1.1}
\centering
\caption{The MWD of polar codes constructed by GA and PW with different code lengths and code rates.}
  \vspace{-0em}
\label{Table_distance_spectrum_polar_codes}
\begin{tabular}{|ccc|m{0.4cm}m{0.8cm}|m{0.4cm}m{0.8cm}|m{0.4cm}m{0.9cm}|m{0.4cm}m{0.9cm}|m{0.4cm}m{1cm}|m{0.4cm}m{1.3cm}|m{0.4cm}m{1.1cm}|}
  \hline
  \multicolumn{3}{|c|}{\multirow{3}{*}{}}&\multicolumn{14}{c|}{$N$}\\
  \cline{4-17}
  &&&\multicolumn{2}{c|}{$256$}&\multicolumn{2}{c|}{$512$}&\multicolumn{2}{c|}{$1024$}&\multicolumn{2}{c|}{$2048$}&\multicolumn{2}{c|}{$4096$}&\multicolumn{2}{c|}{$8192$}&\multicolumn{2}{c|}{$16384$}\\
  \cline{4-17}
  &&&$d_{\min}$&$A_{d_{\min}}$&$d_{\min}$&$A_{d_{\min}}$&$d_{\min}$&$A_{d_{\min}}$&$d_{\min}$&$A_{d_{\min}}$&$d_{\min}$&$A_{d_{\min}}$&$d_{\min}$&$A_{d_{\min}}$&$d_{\min}$&$A_{d_{\min}}$\\

  \hline
  \multicolumn{1}{|c|}{\multirow{30}{*}{$R$}}
  &\multirow{2}{*}{$1/9$}&\multicolumn{1}{|c|}{GA}&32&88&64&4376&64&2608&64&224&128&394848&128&47296&128*&384*\\
  %\cline{3-17}
  \multicolumn{1}{|c|}{}&&\multicolumn{1}{|c|}{PW}&32&88&32&16&64&3120&64&1632&64&704&64&384&64&256\\

  \cline{2-17}
  \multicolumn{1}{|c|}{}&\multirow{2}{*}{$1/8$}&\multicolumn{1}{|c|}{GA}&32&152&32&16&64&8752&64&1376&128&1036896&128&292032&128&128\\
  %\cline{3-17}
  \multicolumn{1}{|c|}{}&&\multicolumn{1}{|c|}{PW}&32&152&32&48&64&6960&64&5216&64&2752&64&1408&64&768\\

  \cline{2-17}
  \multicolumn{1}{|c|}{}&\multirow{2}{*}{$1/7$}&\multicolumn{1}{|c|}{GA}&32&344&32&48&64&19760&64&8288&128&3039840&128&1850560&128&11648\\
  %\cline{3-17}
  \multicolumn{1}{|c|}{}&&\multicolumn{1}{|c|}{PW}&32&280&32&112&32&32&64&14944&64&14528&64&5504&64&4864\\

  \cline{2-17}
  \multicolumn{1}{|c|}{}&\multirow{2}{*}{$1/6$}&\multicolumn{1}{|c|}{GA}&32&920&32&432&64&65328&64&39008&64&1216&128&10958016&128&786816\\
  %\cline{3-17}
  \multicolumn{1}{|c|}{}&&\multicolumn{1}{|c|}{PW}&32&920&32&432&32&96&32&64&64&54464&64&57728&64&45824\\

  \cline{2-17}
  \multicolumn{1}{|c|}{}&\multirow{2}{*}{$1/5$}&\multicolumn{1}{|c|}{GA}&32&2840&32&1840&32&224&64&255584&64&47296&64*&2432*&128&29096320\\
  %\cline{3-17}
  \multicolumn{1}{|c|}{}&&\multicolumn{1}{|c|}{PW}&32&2840&32&2096&32&1376&32&448&32&384&32&256&64&381696\\

  \cline{2-17}
  \multicolumn{1}{|c|}{}&\multirow{2}{*}{$1/4$}&\multicolumn{1}{|c|}{GA}&16&48&32&12592&32&4704&32&64&64&1408192&64&55680&64*&256*\\
  %\cline{3-17}
  \multicolumn{1}{|c|}{}&&\multicolumn{1}{|c|}{PW}&16&48&16&32&32&9312&32&7360&32&3456&32&2816&32&1536\\

  \cline{2-17}
  \multicolumn{1}{|c|}{}&\multirow{2}{*}{$1/3$}&\multicolumn{1}{|c|}{GA}&16&944&16&96&32&161376&32&47296&32&128&64&30026112&64&606976\\
  %\cline{3-17}
  \multicolumn{1}{|c|}{}&&\multicolumn{1}{|c|}{PW}&16&1072&16&608&16&192&16&128&32&158080&32&189184&32&181760\\

  \cline{2-17}
  \multicolumn{1}{|c|}{}&\multirow{2}{*}{$1/2$}&\multicolumn{1}{|c|}{GA}&8&32&16&52832&16&20672&16&896&32&15280512&32&3298048&32&1536\\
  %\cline{3-17}
  \multicolumn{1}{|c|}{}&&\multicolumn{1}{|c|}{PW}&8&96&8&64&16&54464&16&57728&16&45824&16&22016&16&19456\\

  \cline{2-17}
  \multicolumn{1}{|c|}{}&\multirow{2}{*}{$2/3$}&\multicolumn{1}{|c|}{GA}&8&11360&8&5824&8&896&16&3520896&16&2061056&16&230912&16&1024\\
  %\cline{3-17}
  \multicolumn{1}{|c|}{}&&\multicolumn{1}{|c|}{PW}&8&11360&8&11456&8&5504&8&2816&8&3584&8&3072&8&2048\\

  \cline{2-17}
  \multicolumn{1}{|c|}{}&\multirow{2}{*}{$3/4$}&\multicolumn{1}{|c|}{GA}&4&64&8&65728&8&57728&8&23296&8&3584&16&63694336&16&24431616\\
  %\cline{3-17}
  \multicolumn{1}{|c|}{}&&\multicolumn{1}{|c|}{PW}&4&64&8&65728&8&78208&8&90880&8&50688&8&44032&8&38912\\

  \cline{2-17}
  \multicolumn{1}{|c|}{}&\multirow{2}{*}{$4/5$}&\multicolumn{1}{|c|}{GA}&4&448&4&128&8&344448&8&262912&8&108032&8&44032&8&6144\\
  %\cline{3-17}
  \multicolumn{1}{|c|}{}&&\multicolumn{1}{|c|}{PW}&4&448&4&384&4&256&8&508672&8&706048&8&658432&8&366592\\

  \cline{2-17}
  \multicolumn{1}{|c|}{}&\multirow{2}{*}{$5/6$}&\multicolumn{1}{|c|}{GA}&4&1216&4&384&4&256&8&1065728&8&1017344&8&461824&8&186368\\
  %\cline{3-17}
  \multicolumn{1}{|c|}{}&&\multicolumn{1}{|c|}{PW}&4&1216&4&896&4&768&4&512&4&1024&8&2952192&8&9246720\\

  \cline{2-17}
  \multicolumn{1}{|c|}{}&\multirow{2}{*}{$6/7$}&\multicolumn{1}{|c|}{GA}&4&2752&4&1408&4&768&4&512&8&3442176&8&4197376&8&2037760\\
  %\cline{3-17}
  \multicolumn{1}{|c|}{}&&\multicolumn{1}{|c|}{PW}&4&2752&4&2432&4&1792&4&1536&4&1024&4&2048&8&13899776\\

  \cline{2-17}
  \multicolumn{1}{|c|}{}&\multirow{2}{*}{$7/8$}&\multicolumn{1}{|c|}{GA}&4&6848&4&5504&4&2816&4&1536&4&1024&8&11340800&8&15210496\\
  %\cline{3-17}
  \multicolumn{1}{|c|}{}&&\multicolumn{1}{|c|}{PW}&4&6848&4&5504&4&2816&4&1536&4&3072&4&2048&4&4096\\

  \cline{2-17}
  \multicolumn{1}{|c|}{}&\multirow{2}{*}{$8/9$}&\multicolumn{1}{|c|}{GA}&4&12992&4&7552&4&4864&4&3584&4&3072&4&2048&8&36313088\\
  %\cline{3-17}
  \multicolumn{1}{|c|}{}&&\multicolumn{1}{|c|}{PW}&4&12992&4&9600&4&4864&4&5632&4&3072&4&6144&4&4096\\

  \hline
\end{tabular}
\begin{flushleft}
    %\footnotesize{Polar codes except marked with ``*'' are constructed at $E_b/N_0 = 3$dB.\\}
    \footnotesize{Polar codes marked with ``*'' are constructed at $E_b/N_0 = 2.5$dB.}
\end{flushleft}
  \vspace{-0em}
\end{table*}

Since all the codewords of concatenated polar codes with the Hamming weight $d_{\min}$ in the codeword space is on the surface of a sphere, a sphere constraint can also be used to enumerate these codewords with $d_{\min}$. Based on this, an PC-SCEM is proposed to analyze the MWD of concatenated polar codes.

Algorithm \ref{PC-SS} describes the entire procedure of the PC-SCEM. In the method, since $d_{\min}$ determines the sphere constraint, deciding $d_{\min}$ is the first thing to analyze the MWD. However, there are no simple methods to calculate the $d_{\min}$ of concatenated polar codes. Therefore, a greedy method is used to determine $d_{\min}$. We first set the radius of the sphere constraint with a lower bound of $d_{\min}$. Since concatenated polar codes are the subcode of the corresponding polar codes, the minimum Hamming weight of polar codes is the lower bound of the minimum Hamming weight of concatenated polar code. Thus, the radius is first set as
\begin{equation}\label{lower_dmin}
r = \mathop {\min }\limits_{i \in {\cal B}} \left(wt\left({\bf G}_{i}\right)\right).
\end{equation}
Then, considering that polar code is the subcode of RM code and the Hamming weight of the codewords of RM code is even [16, Sec. 4.3,], the Hamming weight of the codewords of concatenated polar codes is even as well. Thus, if no codewords can be enumerated in the sphere constraint (\ref{lower_dmin}), $r$ is added 2 until finding codewords in the sphere constraint and $r$ is the $d_{\min}$ of the concatenated polar code.

To enumerate the codewords in the sphere constraint, all the bits are divided into three types: information bits, frozen bits and parity-check bits. For the information bits and frozen bits, the search process is same as that in the SCEM. For the parity-check bits, they are directly calculated by the previous searched bits according to the corresponding ${\cal Q}_i\left({\bf v}\right)$ such that the codewords searched by the PC-SCEM belong to the MWD of the concatenated polar code.

\section{MWD and Complexity Evaluation}

In this section, we first provide the MWD of polar codes. Then, the optimal CRC polynomial of the CRC-polar concatenated codes and the corresponding MWD are provided. Finally, the complexity comparison between the three proposed methods and the SCL based methods is provided. The improved GA \cite{GA_DAI} and the polarization weight (PW) \cite{PW} are applied to construct polar codes.

\subsection{MWD of Polar Codes}

In this subsection, the MWD of polar codes with different code rates is first provided. Then, we provide the MWD of polar codes with different SNR. Finally, the BLER performance of polar codes decoded by SCL with list size 32 and the corresponding AUB are provided.

Table \ref{Table_distance_spectrum_polar_codes} provides the MWD of polar codes constructed by GA and PW with different code lengths and code rates. Since the MWD of the polar codes constructed by GA changes along with SNR,
the polar codes are constructed at $E_b/N_0 = 3$dB. The polar codes marked with ``*'' are constructed at $E_b/N_0 = 2.5$dB, since $A_{d_{\min}}$ of these polar codes at $E_b/N_0 = 3$dB is so large that the codewords are difficult to be enumerated.
In Table \ref{Table_distance_spectrum_polar_codes}, we can observe that the MWD of polar codes constructed by GA and PW with $N = 256$ is almost the same. Then, the difference of the MWD between GA and PW occurs and becomes larger as the code length increases. Specifically, the MWD of polar codes constructed by GA has larger $d_{\min}$ or less $A_{d_{\min}}$. Based on this, we can explain why GA and PW have almost the same performance for short polar codes, but the performance of GA is better for long polar codes generally.

\begin{table}[t]
\renewcommand\arraystretch{1.1}
\centering
\vspace{-0em}
\caption{The MWD of polar codes constructed by GA in different SNR with code rate $1/2$ and different code lengths.}
\label{distance_spectrum_polar_codes_1024}
\begin{tabular}{|c|m{0.2cm}m{0.2cm}|m{0.2cm}m{0.2cm}|m{0.2cm}m{0.2cm}|m{0.2cm}m{0.2cm}|}
\hline
&\multicolumn{2}{c|}{$(256,128)$}&\multicolumn{2}{c|}{$(512,256)$}&\multicolumn{2}{c|}{$(1024,512)$}&\multicolumn{2}{c|}{$(2048,1024)$}\\
\hline
$\frac{E_b}{N_0}$(dB)&$d_{\min}$&$A_{d_{\min}}$&$d_{\min}$&$A_{d_{\min}}$&$d_{\min}$&$A_{d_{\min}}$&$d_{\min}$&$A_{d_{\min}}$\\
\hline
0.0&8&224&8&64&16&66752&16&86400\\
\hline
0.5&8&224&8&64&16&66752&16&61824\\
\hline
1.0&8&224&8&64&16&54464&16&57728\\
\hline
1.5&8&96&8&64&16&54464&16&33152\\
\hline
2.0&8&96&16&61024&16&45248&16&27008\\
\hline
2.5&8&96&16&58976&16&35008&16&5504\\
\hline
3.0&8&32&16&52832&16&20672&16&896\\
\hline
3.5&8&32&16&44640&16&12992&32&17822912\\
\hline
4.0&16&60720&16&39520&16&5824&32&13382848\\
\hline
4.5&16&60720&16&30816&16&704&32&10843328\\
\hline
\end{tabular}
  \vspace{-0em}
\end{table}
\begin{figure}[t]
\setlength{\abovecaptionskip}{0.cm}
\setlength{\belowcaptionskip}{-0.cm}
  \centering{\includegraphics[scale=0.68]{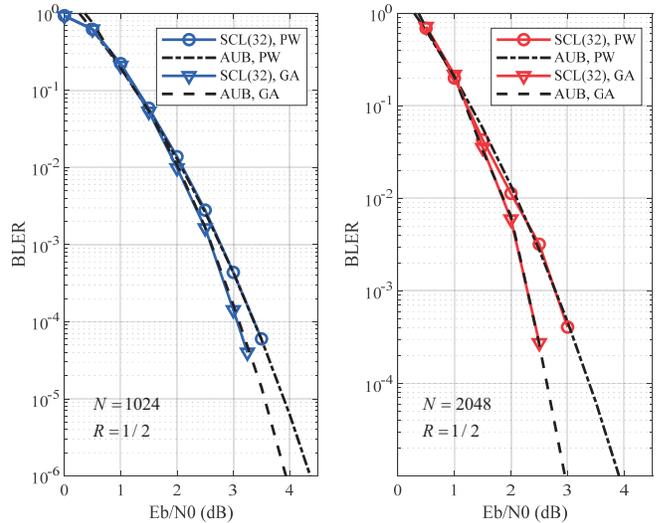}}
  \caption{The BLER performance of polar codes decoded by SCL with list size 32.}\label{N1024_N2048_BLER}
\end{figure}

Table \ref{distance_spectrum_polar_codes_1024} provides the MWD of polar codes with code rate $1/2$ and different code lengths at various SNR. Polar codes are constructed by GA. In Table \ref{distance_spectrum_polar_codes_1024}, we can observe that the MWD of polar codes constructed by GA is variable with the change of SNR.
The reason is that the information set obtained in terms of GA is distinct in different SNR regions.
In addition, the MWD with fixed code length and code rate is also improved as the SNR increases, i.e., larger $d_{\min}$ or less $A_{d_{\min}}$. Thus, no error floor occurs in the performance curve of GA due to the improved MWD. Moreover, compared with the MWD of polar codes constructed by PW (shown in Table I), the polar codes constructed by GA has better AUB in the high SNR region, which leads the performance is better in Fig. \ref{N1024_N2048_BLER}.

Fig. \ref{N1024_N2048_BLER} shows the BLER performance of polar codes with code rate $1/2$ and code length 1024 and 2048. According to Table \ref{Table_distance_spectrum_polar_codes} and Table \ref{distance_spectrum_polar_codes_1024}, the AUB calculated by (\ref{union_bound_a}) is provided. In Fig. \ref{N1024_N2048_BLER}, we can observe that the BLER performance coincides with the corresponding AUB. Thus, the AUB calculated by the MWD can be used to evaluate the BLER performance of polar codes. Then, in the low SNR region, the performance of polar codes constructed by GA is almost the same as that constructed by PW. As the SNR increases, the performance gap between them occurs and becomes larger. The reason is that the MWD of polar codes constructed by GA is improved with the increase of SNR and better than that constructed by PW. Therefore, from the viewpoint of MWD, GA is appropriate for constructing long polar codes rather than PW.

\subsection{MWD of CRC-Polar Concatenated Codes}

In this subsection, we first provide the MWD of CRC-polar concatenated codes with optimal CRC polynomial. Then, the corresponding BLER performance are provided.

Table \ref{distance_spectrum_CRC_polar_codes} provides the MWD of CRC-polar concatenated codes for different code lengths and code rates with optimal CRC and standard CRC. PW is used to construct CRC-polar concatenated codes. The standard CRC polynomials are provided in \cite{CRC8}. By exhausting all the CRC polynomial and analyzing the corresponding MWD of CRC-polar concatenated codes, the optimal CRC is obtained. The optimization principles are 1) maximizing $d_{\min}$ and 2) minimizing $A_{d_{\min}}$ when $d_{\min}$ is identical.

\begin{table}[t]
\renewcommand\arraystretch{1.1}
\centering
\vspace{-0em}
\caption{The MWD of CRC-polar concatenated codes constructed by PW with different code lengths and code rates.}
\label{distance_spectrum_CRC_polar_codes}
\begin{tabular}{|c|c|c|m{0.7cm}|m{0.5cm}|m{0.6cm}|m{0.7cm}|m{0.5cm}|m{0.6cm}|}
\hline
\multirow{2}{*}{$N$}&\multirow{2}{*}{$K_I$}&\multirow{2}{*}{$K_P$}&\multicolumn{3}{c|}{Optimal CRC}&\multicolumn{3}{c|}{Standard CRC}\\
\cline{4-9}
&&&$g(x)$&$d_{\min}$&$A_{d_{\min}}$&$g(x)$&$d_{\min}$&$A_{d_{\min}}$\\
\hline
\multirow{9}{*}{$128$}&\multirow{3}{*}{$32$}&6&0x5B&24&270&0x59&16&12\\
\cline{3-9}
&&8&0x1E7&24&128&0x1D5&16&5\\
\cline{3-9}
&&11&0xD11&24&34&0xCBB&16&3\\
\cline{2-9}
&\multirow{3}{*}{$64$}&6&0x73&12&300&0x59&8&56\\
\cline{3-9}
&&8&0x14D&12&99&0x1D5&8&14\\
\cline{3-9}
&&11&0xD63&12&15&0xCBB&12&147\\
\cline{2-9}
&\multirow{3}{*}{$96$}&6&0x73&6&16&0x59&6&53\\
\cline{3-9}
&&8&0x18D&6&6&0x1D5&4&8\\
\cline{3-9}
&&11&0xECF&8&2453&0xCBB&4&12\\
\hline
\multirow{9}{*}{$256$}&\multirow{3}{*}{$64$}&6&0x79&32&1640&0x59&16&8\\
\cline{3-9}
&&8&0x1F9&32&362&0x1D5&32&758\\
\cline{3-9}
&&11&0x895&32&41&0xCBB&32&136\\
\cline{2-9}
&\multirow{3}{*}{$128$}&6&0x57&16&5853&0x59&12&23\\
\cline{3-9}
&&8&0x1D7&16&1397&0x1D5&12&16\\
\cline{3-9}
&&11&0xC31&16&200&0xCBB&16&553\\
\cline{2-9}
&\multirow{3}{*}{$192$}&6&0x57&8&4647&0x59&8&9494\\
\cline{3-9}
&&8&0x14D&8&1621&0x1D5&8&3521\\
\cline{3-9}
&&11&0xCB9&8&155&0xCBB&8&606\\
\hline
\multirow{9}{*}{$512$}&\multirow{3}{*}{$128$}&6&0x43&32&498&0x59&32&1036\\
\cline{3-9}
&&8&0x1F3&32&95&0x1D5&32&256\\
\cline{3-9}
&&11&0x9A7&32&3&0xCBB&32&32\\
\cline{2-9}
&\multirow{3}{*}{$256$}&6&0x57&16&1912&0x59&16&4344\\
\cline{3-9}
&&8&0x14D&16&362&0x1D5&16&918\\
\cline{3-9}
&&11&0xC23&16&28&0xCBB&16&213\\
\cline{2-9}
&\multirow{3}{*}{$384$}&6&0x43&8&2563&0x59&8&5220\\
\cline{3-9}
&&8&0x187&8&368&0x1D5&8&1193\\
\cline{3-9}
&&11&0xE81&8&6&0xCBB&8&708\\
\hline
\end{tabular}
  \vspace{-0em}
\end{table}

\begin{figure}[t]
\setlength{\abovecaptionskip}{0.cm}
\setlength{\belowcaptionskip}{-0.cm}
  \centering{\includegraphics[scale=0.67]{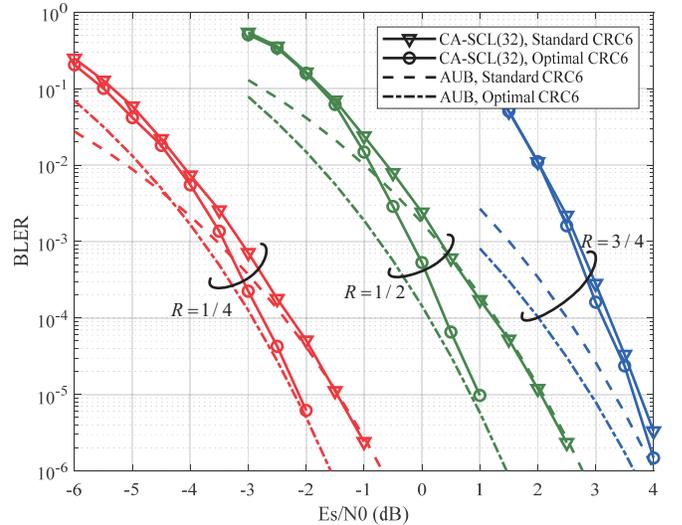}}
  \caption{The BLER performance of CRC-polar concatenated codes with code length 128 and CRC length 6.}\label{N128_CRC_BLER}
\end{figure}

Fig. \ref{N128_CRC_BLER} shows the BLER performance of CRC-polar concatenated codes with the optimal CRC and the standard CRC, where code length is 128 and CRC length is 6. As shown in Table \ref{distance_spectrum_CRC_polar_codes}, the optimal CRC polynomials for code rates $1/4$, $1/2$ and $3/4$ are 0x5B, 0x73 and 0x73, respectively, and the standard CRC polynomial is 0x59.
In Fig. \ref{N128_CRC_BLER}, the BLER performance is close to the AUB in the high SNR region. Then, since the CRC-polar concatenated codes with the optimal CRC has better MWD, the performance is better in the medium to high SNR regions.

\begin{figure}[t]
\setlength{\abovecaptionskip}{0.cm}
\setlength{\belowcaptionskip}{-0.cm}
  \centering{\includegraphics[scale=0.67]{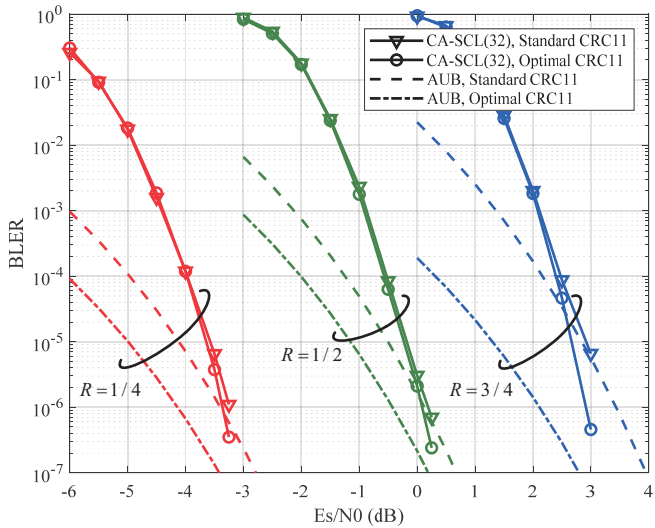}}
  \caption{The BLER performance of CRC-polar concatenated codes with code length 512 and CRC length 11.}\label{N512_CRC_BLER}
\end{figure}

Fig. \ref{N512_CRC_BLER} illustrates the BLER performance of CRC-polar concatenated codes with code length 512 and 11-bit CRC. As shown in Table \ref{distance_spectrum_CRC_polar_codes}, the optimal CRC polynomials for code rates $1/4$, $1/2$ and $3/4$ are 0x9A7, 0xC23 and 0xE81, respectively, and the standard CRC polynomial is 0xCBB.
Similarly to Fig. \ref{N128_CRC_BLER}, the BLER performance is also close to the AUB and the performance of the optimal CRC is better in the high SNR region.

\subsection{Complexity Evaluation}

In this subsection, we provide the complexity comparison between the proposed methods and the SCL based methods.

\begin{figure}[t]
\setlength{\abovecaptionskip}{0.cm}
\setlength{\belowcaptionskip}{-0.cm}
  \centering{\includegraphics[scale=0.67]{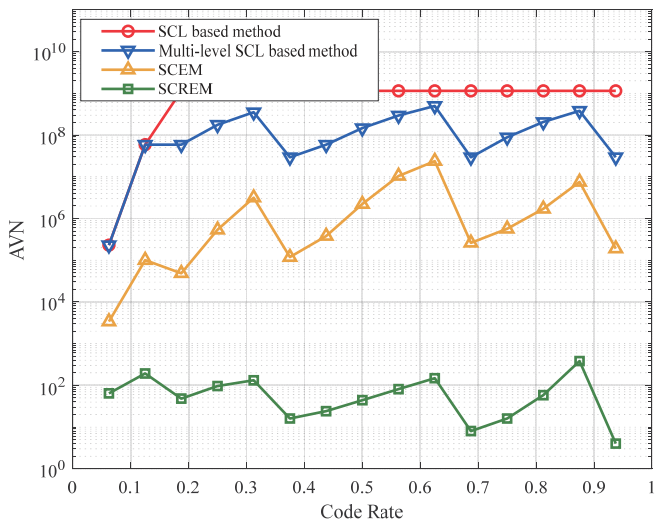}}
  \caption{The complexity of the SCEM, the SCREM and the SCL based methods with $N = 128$.}\label{N128_Complexity}
\end{figure}

Fig. \ref{N128_Complexity} illustrates the complexities of the SCEM, the SCREM and the SCL based methods with code length 128 and different code rates. Considering the MWD of polar codes constructed by GA changes along with SNR, PW is used in Fig. \ref{N128_Complexity}. The complexity of SCEM and SCREM is counted by the average visited nodes (AVN). The AVN of enumerating all the codewords is $2^KN\log N$, which is the upper bound of the complexity of enumerating MWD. The AVN of SCL method \cite{dsliu} with list size $L_1$ are $\min\left(2^KN\log N,~ L_1N\log N\right)$. The AVN of multi-level SCL method \cite{CRCdesign} with list size $L_2$ and level number $M$ are $\min\left(2^KN\log N,~ ML_2N\log N\right)$. $L_1$ and $L_2$ used in \cite{dsliu} and \cite{CRCdesign} are 1280000 and 32768, respectively, and $M$ is the number of row with the row weight $d_{\min}$ in the generator matrix of polar codes.

In Fig. \ref{N128_Complexity}, the complexity of SCEM is lower than those of the SCL based methods, since the sphere constraint can prune the search tree to reduce the redundant search.
Specifically, the AVN of the proposed SCEM achieves 3 to 4 magnitude reduction compared with both the SCL method and the multi-level SCL method.
Furthermore, due to the recursive structure of SCREM, its complexity is lower than the SCEM. In comparison to the SCEM, the SCL method and the multi-level SCL method, the SCREM can achieve up to $10^5$, $10^8$, and $10^7$ times complexity reduction.

%Moreover, both the SCL method and the multi-level SCL method are breadth-first searching, which leads that $L_1$ and $ML_2$ survival paths need to be stored in the memory, respectively.
%Thus, they both have high memory overheads and difficult to be implemented in a practice memory-constrained computer.
%However, the proposed SCREM and SCEM have negligible memory overhead, due to only one path stored in the memory.
%Furthermore, some codewords with the minimum Hamming weight may be lost in the SCL based methods when the list size is not sufficiently large. On the contrary, the proposed SCREM and SCEM can enumerate all the codewords belonging to the MWD.

\section{Conclusion}

In this paper, we propose three methods to analyze the MWD of polar codes. The SCEM is first proposed to exactly enumerate all the codewords belonging to the MWD, which exploits the distance property that all the codewords with the identical Hamming weight are distributed on a spherical shell. Then, based on the SCEM and the Plotkin's construction, we propose SCREM to recursively analyze the MWD with lower complexity. Finally, the PC-SCEM is proposed by introducing the parity-check equations and the sphere constraint to analyze the MWD of concatenated polar codes. The experimental results illustrate that the complexities of the proposed SCEM and SCREM with code length 128 are  lower than those of the SCL based methods.

\ifCLASSOPTIONcaptionsoff
  \newpage
\fi

\begin{appendix}
  %\section{The Proof of Lemma \ref{lemma4_distance_spectrum_d1_is_d2}  }
According to (\ref{Plotkin¡¯s construction}), a codeword $\bf c$ with minimum Hamming weight $d_{\min}$ can be expressed as
\begin{equation}\label{c_d1_d2_dmin}
wt\left({\bf c}\right) =  wt\left({\bf c}' + {\bf c}''\right) + wt\left( {\bf c}''\right) = d_{\min}.
\end{equation}
Then, $d'_{\min}$ and $d''_{\min}$ are divided into three cases by Lemma \ref{lemma3_d2_d1}. In the proof of each case, classified discussion is used.
\begin{enumerate}[1)]
\item When $d'_{\min}=d_{\min}$ and $d''_{\min}=d_{\min}$, $\cal T$ is obtained as follows.
    \begin{enumerate}[a)]
    \item Supposing $wt\left( {\bf c}''\right) = 0$, $wt\left({\bf c}\right)$ is simplified as
        \begin{equation}
        wt\left({\bf c}\right) = wt\left({\bf c}'\right) = d_{\min}.
        \end{equation}
        Thus, $\forall {\bf c}' \in {\cal T}'$ makes $wt\left({\bf c}\right)$ is $d_{\min}$.
    \item Supposing $wt\left( {\bf c}''\right) = d_{\min}$, similarly,  $wt\left({\bf c}\right)$ is simplified as
        \begin{equation}
        wt\left({\bf c}' + {\bf c}''\right) = 0.
        \end{equation}
        Hence, we have ${\bf c}'' = {\bf c}'$.
        Furthermore, according to Lemma \ref{lemma1_subcode}, we have ${\cal T}' \subset {\cal T}''$. Therefore, for $\forall {\bf c}' \in {\cal T}'$, $\exists {\bf c}'' \in {\cal T}''$ makes $wt\left({\bf c}' + {\bf c}''\right) = 0$, i.e., ${\bf c}'' = {\bf c}'$. Thus, $\left({\bf 0}, {\bf c}'\right)$, ${\bf c}' \in {\cal T}'$, is the codeword of $\cal C$ and its Hamming weight is $d_{\min}$.
    \item Supposing $wt\left( {\bf c}''\right) > d_{\min}$, it is clear that $wt\left({\bf c}\right) > d_{\min}$.
    \end{enumerate}
    In conclusion, ${\cal T} = {\cal T}_1 \cup {\cal T}_2$, where ${\cal T}_1 = \left\{\left({\bf c}', {\bf 0}\right) | {\bf c}' \in {\cal T}'\right\}$ and ${\cal T}_2 = \left\{\left({\bf 0}, {\bf c}'\right) | {\bf c}' \in {\cal T}'\right\}$.
\item When $d'_{\min}=d_{\min}$ and $d''_{\min}=\frac{d_{\min}}{2}$, $\cal T$ is obtained as follows.
    \begin{enumerate}[a)]
    \item Supposing $wt\left( {\bf c}''\right) = 0$, $\forall {\bf c}' \in {\cal T}'$ makes $wt\left({\bf c}\right)$ is $d_{\min}$.
    \item Supposing $wt\left( {\bf c}''\right) = \frac{d_{\min}}{2}$ and $wt\left( {\bf c}'\right) = 0$, obviously, $\left({\bf c}'', {\bf c}''\right)$ is the codeword of $\cal C$ and its Hamming weight is $d_{\min}$.
    \item Supposing $wt\left( {\bf c}''\right) = \frac{d_{\min}}{2}$ and $wt\left( {\bf c}'\right) = d_{\min}$, to obtain the codeword $\bf c$ with $d_{\min}$, all the ${\bf c}' \in {\cal T}'$ and ${\bf c}'' \in {\cal T}''$ are enumerated to satisfy
        \begin{equation}\label{no_codeword}
        wt\left({\bf c}' + {\bf c}''\right) = \frac{d_{\min}}{2}.
        \end{equation}
    \item Supposing $wt\left( {\bf c}''\right) = \frac{d_{\min}}{2}$ and $wt\left( {\bf c}'\right) > d_{\min}$, it is clear that $wt\left({\bf c}' + {\bf c}''\right) > \frac{d_{\min}}{2}$.
        Thus, $wt\left({\bf c}\right) > d_{\min}$.
    \item Supposing $\frac{d_{\min}}{2} < wt\left( {\bf c}''\right) < d_{\min}$, in order to make $wt\left({\bf c}\right)$ is $d_{\min}$, we have
        \begin{equation}\label{no_codeword_1}
        0< wt\left({\bf c}' + {\bf c}''\right) < \frac{d_{\min}}{2}.
        \end{equation}
        Then, according to Lemma \ref{lemma1_subcode}, ${\bf c}'$ is the codeword of ${\cal C}''$. Thus, ${\bf c}' + {\bf c}''$ is also the codeword of ${\cal C}''$. However, since $d''_{\min}$ is $\frac{d_{\min}}{2}$, no codeword of ${\cal C}''$ can satisfy (\ref{no_codeword_1}). Therefore, in this case, no codeword of $\cal C$ with $d_{\min}$ can be found.
    \item Supposing $wt\left( {\bf c}''\right) = d_{\min}$, $wt\left({\bf c}\right)$ is simplified as
        \begin{equation}
        wt\left({\bf c}' + {\bf c}''\right) = 0.
        \end{equation}
        According to the 1)-b) of the proof of Lemma \ref{lemma4_distance_spectrum_d1_is_d2},
        $\left({\bf 0}, {\bf c}'\right)$, ${\bf c}' \in {\cal T}'$, is the codeword of $\cal C$ and its Hamming weight is $d_{\min}$.
    \item Supposing $wt\left( {\bf c}''\right) > d_{\min}$, obviously, $wt\left({\bf c}\right) > d_{\min}$.
    \end{enumerate}
    In conclusion, ${\cal T} = {\cal T}_1 \cup {\cal T}_2 \cup {\cal T}_3 \cup {\cal T}_4$, where ${\cal T}_3 = \left\{ \left( {\bf c}'',{\bf c}''\right)|{\bf c}''\in{\cal T}'' \right\}$ and ${\cal T}_4 = \{ \left( {\bf c}' \oplus {\bf c}'',{\bf c}''\right)|{\bf c}'\in{\cal T}', {\bf c}''\in{\cal T}'', wt({\bf c}' \oplus {\bf c}'') = \frac{d_{\min}}{2} \}$.

\item When $d'_{\min}>d_{\min}$ and $d''_{\min}=\frac{d_{\min}}{2}$, $\cal T$ is obtained as follows.
    \begin{enumerate}[a)]
    \item Supposing $wt\left( {\bf c}''\right) = 0$, it is clear that $wt\left({\bf c}\right) > d_{\min}$.
    \item Supposing $wt\left( {\bf c}''\right) = \frac{d_{\min}}{2}$ and $wt\left( {\bf c}'\right) = 0$, obviously, $\left({\bf c}'', {\bf c}''\right)$ is the codeword of $\cal C$ and its Hamming weight is $d_{\min}$.
    \item Supposing $wt\left( {\bf c}''\right) = \frac{d_{\min}}{2}$ and $wt\left( {\bf c}'\right) > d_{\min}$, we have
        \begin{equation}
        wt\left({\bf c}' + {\bf c}''\right) > \frac{d_{\min}}{2}.
        \end{equation}
        Thus, $wt\left({\bf c}\right) > d_{\min}$.
    \item Supposing $ \frac{d_{\min}}{2} < wt\left( {\bf c}''\right) < d_{\min}$, in order to make $wt\left({\bf c}\right)$ is $d_{\min}$, we have
        \begin{equation}\label{no_codeword}
        0 < wt\left({\bf c}' + {\bf c}''\right) < \frac{d_{\min}}{2}.
        \end{equation}
        Then, according to the 2)-e) of the proof of Lemma \ref{lemma4_distance_spectrum_d1_is_d2}, no codeword of $\cal C$ with $d_{\min}$ can be found in this case.
    \item Supposing $wt\left( {\bf c}''\right) = d_{\min}$, we have $wt\left({\bf c}' + {\bf c}''\right) > 0$. Thus, $wt\left({\bf c}\right) > d_{\min}$.
    \item Supposing $wt\left( {\bf c}''\right) > d_{\min}$, obviously, $wt\left({\bf c}\right) > d_{\min}$.
    \end{enumerate}
    In conclusion, ${\cal T}_3 = \left\{ \left( {\bf c}'',{\bf c}''\right)|{\bf c}''\in{\cal T}'' \right\}$.
\end{enumerate}
From the above, Lemma \ref{lemma4_distance_spectrum_d1_is_d2} has been proved.
\end{appendix}


\begin{thebibliography}{99}

\bibitem{arikan}
E. Ar{\i}kan, ``Channel polarization: A method for constructing capacity achieving codes for symmetric binary-input memoryless channels,'' \emph{IEEE Trans. Inf. Theory}, vol. 55, no. 7, pp. 3051--3073, Jul. 2009.

\bibitem{talvardyscl}
I. Tal and A. Vardy, ``List decoding of polar codes,'' \emph{IEEE Trans. Inf. Theory}, vol. 61, no. 5, pp. 2213--2226, May. 2015.

\bibitem{niuscl}
K. Chen, K. Niu, and J. R. Lin, ``List successive cancellation decoding
of polar codes,'' \emph{Electron. Lett.}, vol. 48, no. 9, pp. 500--501, 2012.

\bibitem{SCS}
K. Chen, K. Niu and J. Lin, ``Improved Successive Cancellation Decoding of Polar Codes,'' \emph{IEEE Trans. Commun.}, vol. 61, no. 8, pp. 3100--3107, Aug. 2013.

\bibitem{niu_CASCL}
K. Niu and K. Chen, ``CRC-aided decoding of polar codes,'' \emph{IEEE Commun. Lett.}, vol. 16, no. 10, pp. 1668--1671, Oct. 2012.

\bibitem{3GPP_5G_polar}
3$^\text{rd}$ Generation Partnership Project (3GPP), ``Multiplexing and channel coding,'' \emph{3GPP TS 38.212 V15.0.0}, 2017.

\bibitem{FAR_DL}
J. Dai, J. Gao and K. Niu, ``Learning to Mitigate the FAR in Polar Code Blind Detection,'' \emph{IEEE Wireless Commun. Lett.}, early access.

\bibitem{ADSCL}
B. Li, H. Shen and D. Tse, ``An Adaptive Successive Cancellation List Decoder for Polar Codes with Cyclic Redundancy Check,'' \emph{IEEE Commun. Lett.}, vol. 16, no. 12, pp. 2044--2047, December 2012.

\bibitem{LinShu}
S. Lin and D. J. Costello, ``Error Control Coding (2nd ed.),'' PrenticeHall, Inc., 2004.

%\bibitem{dmin}
%A. Eslami and H. Pishro-Nik, ``On Finite-Length Performance of Polar Codes: Stopping Sets, Error Floor, and Concatenated Design,'' \emph{IEEE Trans. Commun.}, vol. 61, no. 3, pp. 919--929, March 2013.

\bibitem{dsliu}
Z. Liu, K. Chen, K. Niu and Z. He, ``Distance spectrum analysis of polar codes,'' in \emph{Proc. 2014 IEEE WCNC}, Istanbul, 2014, pp. 490--495.

\bibitem{CRCdesign}
Q. Zhang, A. Liu, X. Pan and K. Pan, ``CRC Code Design for List Decoding of Polar Codes,'' \emph{IEEE Commun. Lett.}, vol. 21, no. 6, pp. 1229--1232, June 2017.

\bibitem{SCPC}
E. Ar{\i}kan, ``Serially Concatenated Polar Codes,'' \emph{IEEE Access}, vol. 6, pp. 64549--64555, 2018.

\bibitem{CRCds}
G. Ricciutelli, T. Jerkovits, M. Baldi, F. Chiaraluce and G. Liva, ``Analysis of the Block Error Probability of Concatenated Polar Code Ensembles,'' \emph{IEEE Trans. Commun.}, vol. 67, no. 9, pp. 5953--5962, Sep. 2019.

\bibitem{SD}
S. Kahraman and M. E. Celebi, ``Code based efficient maximum-likelihood decoding of short polar codes,'' in \emph{Proc. 2012 IEEE Int. Symp. Inf. Theory}, Cambridge, MA, pp. 1967--1971, 2012.

\bibitem{niu_SD}
K. Niu, K. Chen and J. Lin, ``Low-Complexity Sphere Decoding of Polar Codes Based on Optimum Path Metric,'' \emph{IEEE Commun. Lett.}, vol. 18, no. 2, pp. 332--335, Feb. 2014.

\bibitem{efficient_SD}
J. Guo and A. Guill¨¦n i F¨¤bregas, ``Efficient sphere decoding of polar codes,'' in \emph{Proc. 2015 IEEE Int. Symp. Inf. Theory}, Hong Kong, pp. 236--240, 2015.

\bibitem{CASD}
J. Piao, J. Dai and K. Niu, ``CRC-Aided Sphere Decoding for Short Polar Codes,'' \emph{IEEE Commun. Lett.}, vol. 23, no. 2, pp. 210--213, Feb. 2019.

\bibitem{Plotkin}
S. Ejaz, F. Yang, T. H. Soliman, ``Network polar coded cooperation with joint sc decoding'', \emph{Electron. Lett.}, vol. 51, no. 9, pp. 695--697, 2015.

\bibitem{PO}
C. Sch¨¹rch, ``A partial order for the synthesized channels of a polar code,'' in \emph{Proc. 2016 IEEE Int. Symp. Inf. Theory}, Barcelona, pp. 220--224, 2016.

%\bibitem{GA_Trifonov}
%P. Trifonov, ``Efficient design and decoding of polar codes,'' \emph{IEEE Trans. Commun.}, vol. 60, no. 11, pp. 3221--3227, Nov. 2012.

\bibitem{GA_DAI}
J. Dai, K. Niu, Z. Si, C. Dong and J. Lin, ``Does Gaussian Approximation Work Well for the Long-Length Polar Code Construction?,'' \emph{IEEE Access}, vol. 5, pp. 7950--7963, 2017.

\bibitem{PW}
G. He et al., ``Beta-Expansion: A Theoretical Framework for Fast and Recursive Construction of Polar Codes,'' in \emph{Proc. 2017 IEEE Global Communications Conference}, Singapore, pp. 1--6, 2017.

\bibitem{CRC8}
P. Koopman and T. Chakravarty, ``Cyclic redundancy code (CRC) polynomial selection for embedded networks,'' in \emph{International Conference on Dependable Systems and Networks}, Florence, Italy, 2004, pp. 145--154, 2004.


\end{thebibliography}
\end{document}